\documentclass[12pt]{article}
\newcommand{\correction}[1]{#1}

\title{WKB approximation in
deformed space with minimal length}
\author{
T.V. Fityo\footnote{E-mail: fityo@ktf.franko.lviv.ua},\ \ I.O.
Vakarchuk\footnote{E-mail: chair@ktf.franko.lviv.ua}\ \ and V.M.
Tkachuk\footnote{E-mail: tkachuk@ktf.franko.lviv.ua}
\\
  {\small Chair of Theoretical Physics, Ivan Franko National University of Lviv,}\\
{\small 12 Drahomanov St., Lviv, UA-79005, Ukraine
         }}

\begin{document}
\maketitle

\abstract{\correction{The} WKB approximation for deformed space
with minimal length is considered. \correction{The}
Bohr-Sommerfeld quantization rule is obtained. A new interesting
feature in presence of deformation is that \correction{the} WKB
approximation is valid for intermediate quantum numbers and can be
invalid for small as well as very large quantum numbers.
\correction{The} correctness of the rule is verified by comparing
obtained results with exact expressions for corresponding
spectra.}

\vspace{.5cm} Keywords: deformed Heisenberg algebra, WKB
approximation, Bohr-Sommerfeld quantization rule.

PACS numbers: 02.40.Gh, %Noncommutative geometry
03.65.Sq,    %Semiclassical theories and applications
%31.15.Gy     %Semiclassical methods in Calculations and mathematical techniques in atomic and molecular physics (excluding electron correlation calculations)

\section{Introduction}
Quantum mechanics with modification of the usual canonical
commutation relations has being investigated intensively lately.
Such works are motivated by several independent lines of
investigations in string theory and quantum gravity, which suggest
the existence of finite lower bound to the possible resolution of
length $\Delta X$ \cite{Gross88, Maggiore93, Witten96}.

A lot of attention was payed to the following deformed commutation
relation \cite{Kempf95, Kempf97, Kempf00, Detournay02}:
\begin{equation}\label{in1}
[X,P]=i\hbar(1+\beta P^2)
\end{equation}
and it was shown that it implies existence of minimal resolution
length $\Delta X=\sqrt{\langle(\Delta X)^2 \rangle}\ge
\hbar\sqrt\beta$ \cite{Kempf95}, i.e., there is no possibility to
measure coordinate $X$ with accuracy larger than $\Delta X$. If
someone puts $\beta=0$ the usual Heisenberg algebra be obtained.

The use of the deformed commutation relations (\ref{in1}) brings
new difficulties in solving the quantum problems. As far as we
know there are only few problems for which spectra have been found
exactly. They are 1 dimensional oscillator \cite{Kempf95}, D
dimensional isotropic harmonic oscillator \cite{Chang02}, 3
dimensional relativistic Dirac oscillator \cite{Quesne05}, 1D
Coulomb potential \cite{Fityo05}. Note, that in 1 dimensional case
the harmonic oscillator problem has been solved exactly
\cite{Quesne03, Quesne04} for more general deformation leading to
nonzero uncertainties in both position and momentum.

Difficulties of obtaining exact solution of quantum problems
originates development of perturbation techniques \cite{Kempf97,
Kempf97b, Brau99, Benczik05} and numerical calculus
\cite{Benczik05} in presence of deformation. In our recent work
\cite{Fityo05} we derived exact expression for the spectrum of 1D
Coulomb potential and we also obtained the same result with the
help of Bohr-Sommerfeld quantization rule. The use of the rule was
intuitional, in this paper we derive this rule rigidly and analyze
its applicability.

The present paper is organized as follow\correction{s}. In the
second section, WKB approximation is extended for deformed
commutation relation, Bohr-Sommerfeld quantization rule is
obtained and its applicability is discussed. In the third section,
several 1D examples are analyzed and obtained spectra are compared
to \correction{the} exact results. In the fourth section, we show
that the quantization rule can be applied to 3D problems with
radial symmetry. And finally the paper is ended with concluding
remarks.

\section{WKB approximation}

The 1D Schr\"odinger equation in deformed space reads
\begin{equation}\label{h1}
\left[\frac{P^2}{2m}+U(X)\right]\psi=E\psi,
\end{equation}
where the first term in the brackets describes kinetic energy of
the system and the second term describes the potential one. In
deformed space coordinate and momentum operators satisfy the
following commutation relation
\begin{equation}\label{h2}
[X,P]=i\hbar f(P),
\end{equation}
where $f(P)$ is an arbitrary function of $P$, in general $f(P)\ne
1$. We require that $f(P)$ is an even function. This provides
invariance of relation (\ref{h2}) with respect to the reflection
$X\to -X$, $P\to -P$. Here we consider general form of
deformation, particular form of deformation (\ref{in1}) will be
analyzed in more details at the end of the section.

In order to study the semiclassical approximation we use the
so-called quasi-coordinate representation
\begin{equation}\label{h3}
X=x,\qquad P=P(p),\qquad p=-i\hbar\frac d{dx},
\end{equation}
From definitions (\ref{h2}) and (\ref{h3}) we obtain
\begin{equation}\label{h3p}
\frac{dP(p)}{dp}=f(P)
\end{equation}
and $P(p)$ is an odd function.

Let us express wave function in the following form
\begin{equation}\label{h4}
\psi(x)=\exp\left[\frac{i}{\hbar}S(x) \right]\
\end{equation}
then in linear approximation over $\hbar$
\begin{equation}\label{h5}
P^2\psi(x)=\left[ P^2(S'(x))-\frac{i\hbar}2
[P^2(S'(x))]''S''(x)+\dots \right]\psi(x),
\end{equation}
here prime denotes derivative with respect to the argument of the
function. Expanding $S(x)$ in power series over $\hbar$
\begin{equation}\label{h6}
S(x)=S_0(x)+\frac{\hbar}iS_1(x)+\dots
\end{equation}
we obtain the following set of equations for $S$:
\begin{eqnarray}
\frac{P^2(S_0'(x))}{2m}+U(x)=E,\\
\frac{[P^2(S_0'(x))]'}{2m}\frac{\hbar}iS_1'(x)-\frac{i\hbar}{4m}
[P^2(S_0'(x))]''S_0''(x)=0.
\end{eqnarray}
Solutions of these equations read
\begin{equation}\label{h9}
S_0(x)=\int^xp\left(\pm\sqrt{2m(E-U(x))}\right)dx=
\pm\int^xp(P)dx,
\end{equation}
\begin{equation}\label{h10}
S_1(x)=-\frac12\ln\left|{\left[P^2(S_0'(x))\right]'}\right|=-\frac12
\ln\left|2P(x)f(P)\right|,
\end{equation}
where $P=\sqrt{2m(E-U(x))}$ is a function of $x$; $p(P)$ is an
inverse function to $P(p)$, it is taken into account that $p(P)$
is also the odd function. Here we omit constants of integration
since they are taken into account in the final expression for the
wave function:
\begin{eqnarray}\label{h11}
\psi(x)=\frac{1}{\sqrt{\left|P f(P)\right|}} \left( C_1
\exp\left[\frac{i}{\hbar} \int^xpdx\right]+
 C_2 \exp\left[ -\frac{i}{\hbar} \int^xpdx \right]\right).
\end{eqnarray}

To obtain an expression of Bohr-Sommerfeld quantization rule we
have to analyze the behavior of wave function (\ref{h11}) at the
infinities and consider matching conditions near the turning
points. For bound states this analysis leads to the following
Bohr-Sommerfeld quantization condition
\begin{equation}\label{h12}
\int\limits^{x_2}_{x_1} pdx = \pi\hbar(n+\delta), \quad
n=0,1,2,\dots
\end{equation}
where $x_1$ and $x_2$ are turning points satisfying equation
$U(x)=E$, $\delta$ depends on boundary conditions and properties
of $P(x)$. If potential $U(x)$ is a smooth function and if
$f(0)\ne0$ then $\delta=1/2$.

Note, that new small operators $x$ and $p$ satisfy canonical
commutation relation, therefore condition (\ref{h12}) is expected
and was used in our previous work \cite{Fityo05} as an evident
one. In that paper the following recipe to find a spectrum with
the help of Bohr-Sommerfeld quantization rule was applied: we
rewrite the problem in \correction{the} small operators $x$ and
$p$ which satisfy $[x,p]=i\hbar$, then from equation $H(x,p)=E$ we
find out $p=p(x,E)$ and use Bohr-Sommerfeld quantization rule
(\ref{h12}).

The quantization rule (\ref{h12}) can be rewritten in more
convenient form using the following transformations
\begin{equation}\label{h15}
\oint pdx=-\oint xdp=-\oint xdP\frac{dp}{dP}.
\end{equation}
Then taking into account expressions (\ref{h3}) and (\ref{h3p}),
we obtain equivalent form of \correction{the} Bohr-Som\-mer\-feld
quantization rule
\begin{equation}\label{h12s}
-\oint\frac{XdP}{f(P)}=2\pi\hbar(n+\delta).
\end{equation}
This rule does not demand knowing of representation of initial
operators $X$ and $P$ in terms of canonical operators $x$ and $p$
and can be applied to an eigenvalue problem at once.

The WKB approximation is valid if the second term of expansion
(\ref{h5}) is much less than the first term. %After simple
%algebraic transformations we obtain
\correction{Namely, it is valid if
\begin{equation}
P^2\gg \frac{\hbar}2 \left|[P^2(S'(x))]''S''(x)\right|\approx
\frac{\hbar}2\left| \frac{d^2P^2}{dp^2} \frac{dp}{dx} \right|,
\end{equation}
here we substitute $S'(x)$ with $p=S_0'(x)$. This substitution is
correct in linear approximation over $\hbar$. Then, using the fact
that $\frac{d^2P^2}{dp^2}=\frac d{dp}(2Pf(P))$, we obtain}
\begin{equation}\label{h14}
P^2\gg \hbar\left|\frac{d}{dx}Pf(P)\right|.
\end{equation}
In undeformed case it is considered that WKB approximation is
valid for large quantum number $n$. In case of deformation,
condition (\ref{h14}) can be violated for large values of quantum
numbers. We analyze this violation in more details for special
case of deformation
\begin{equation}\label{i1}
f(P)=(1+\beta P^2).
\end{equation}
This case corresponds to
\begin{equation}\label{i2}
P(p)=\frac1{\sqrt\beta}\tan\sqrt\beta p,\qquad p(P)
=\frac1{\sqrt\beta} \arctan\sqrt\beta P.
\end{equation}
Such a deformation (\ref{i1}-\ref{i2}) for smooth potential energy
$U(x)$ gives $\delta=1/2$.

Condition (\ref{h14}) reads
\begin{equation}\label{i4}
P^2\gg \hbar (1+3\beta P^2)\left|\frac{dP}{dx}\right|.
\end{equation}

For small momentum ($\beta P^2\ll1$) we obtain usual condition for
validity of WKB approximation \cite{Landau}:
\begin{equation}\label{i5}
\hbar\left|\frac{d(1/P)}{dx}\right|\ll1.
\end{equation}
For large $P$ we obtain that the following inequality must hold:
\begin{equation}\label{i6}
3\hbar \beta \left|\frac{dP}{dx}\right|\ll1.
\end{equation}

Let us use rough approximation
\begin{equation}\label{i6e}
\left|\frac{dP}{dx}\right|\approx\frac{P}{a}=\frac{2\pi\hbar}{\lambda
a},
\end{equation}
where $a$ is a characteristic size of the system being about
$x_2-x_1$, $\lambda$ is a wave length corresponding to momentum
$P$. It allows with the use of formulae (\ref{i5}), (\ref{i6}) to
estimate ranges in which WKB approximation is valid:
\begin{equation}\label{i7}
a\gg\lambda\gg\frac{\Delta X^2}{a},
\end{equation}
where $\Delta X=\hbar\sqrt\beta$ is a minimal resolution length.
It is interesting to note that if $a\approx \Delta X$ then WKB
approximation is not valid for any momentum value.
\correction{This result is not an unexpected one because a
characteristic size of the system must be larger than minimal
resolution length, otherwise all mathematics and physics become
meaningless.}

\section{1D examples}

\subsection{Harmonic oscillator}\label{ho}

Hamiltonian of the system is
\begin{equation}\label{g1}
H=P^2+X^2,
\end{equation}
here and below we put $m=1/2$, $\omega=2$ and $\hbar=1$ for the
sake of simplicity.

From $H(P,X)=E$ we obtain
\begin{equation}\label{g4}
X=\sqrt{E-P^2}
\end{equation}
and Bohr-Sommerfeld quantization condition (\ref{h12s}) gives
\begin{equation}\label{g5}
2\int\limits^{\sqrt E}_{-\sqrt E} \frac{\sqrt{E-P^2}}{1+\beta
P^2}dP=\frac{2\pi}\beta\left( \sqrt{1+\beta E} -1
\right)=2\pi(n+1/2).
\end{equation}
From the last equation we obtain
\begin{equation}\label{g6}
E_n=(2n+1)+\beta\left(n^2+n+\frac14\right).
\end{equation}

\correction{The} exact result obtained by Kempf and collaborators
\cite{Kempf95} is
\begin{equation}\label{g7}
E_n=(2n+1)\left( \frac\beta2+\sqrt{1+\frac{\beta^2}4}
\right)+\beta n^2\approx (2n+1)+\beta\left(n^2+n+\frac12\right)+
O(\beta^2).
\end{equation}
As one can see, results presented by formulae (\ref{g6}) and
(\ref{g7}) asymptotically coincide for large $n$.

\correction{The} inequality (\ref{i7}) for harmonic oscillator
reads
\begin{equation}\label{g7a}
\sqrt n\gg\frac1{\sqrt n}\gg\frac\beta{\sqrt n}.
\end{equation}
So, if $\beta\ll 1$ it simplifies to usual condition of WKB
approximation applicability
$$n\gg1.$$

\subsection{Anharmonic oscillator} \label{aho}

\correction{The} Hamiltonian reads
\begin{equation}\label{g23}
H=P^2+\gamma^N X^N,
\end{equation}
here $N$ is an even integer.

Then from $H(P,X)=E$ we obtain
\begin{equation}\label{g24}
X=\frac1\gamma(E-P^2)^{1/N}
\end{equation}
and
\begin{equation}\label{g25}
2\int\limits^{\sqrt E}_{-\sqrt E}\frac{XdP}{1+\beta
P^2}=2\pi(n+1/2).
\end{equation}
We calculate the last integral in linear approximation over
$\beta$ and obtain that
\begin{equation}\label{g26}
E_n=E_n^0\left(1+\frac{2\beta}{(1+2/N)(3+2/N)}E_n^0\right),
\end{equation}
where $E_n^0$ denotes energy levels obtained using Bohr-Sommerfeld
quantization rule for $\beta=0$ and they read
\begin{equation}\label{g27}
E_n^0=\left[\pi \frac{\Gamma(3/2+1/N)} {\Gamma(1/2)\Gamma(1+1/N)}
\gamma(n+1/2) \right]^{\frac{2N}{2+N}}.
\end{equation}
If $N=2$ and $\gamma=1$ we reproduce result (\ref{g6}) obtained
for 1D harmonic oscillator.

In limit $N\to\infty$ we obtain system which is equivalent to
infinitely high potential well. For this case (\ref{g26}) gives
\begin{equation}\label{g28b}
E_n=\left(\frac{\pi\gamma(n+\frac12)}{2}\right)^2+\frac23\beta
\left(\frac{\pi \gamma(n+\frac12)}{2}\right)^4,
\end{equation}
Here $2/\gamma$ stands for the well width. Previously the problem
has been considered in \cite{Detournay02} for general form of
deformation function $f(P)$ and in linear approximation over
$\beta$ their approach for $f(P)=1+\beta P^2$ gives
\begin{equation}\label{g28}
E_n=\left(\frac{\pi n\gamma}{2}\right)^2+\frac23\beta
\left(\frac{\pi n\gamma}{2}\right)^4,
\end{equation}
The difference between formulae (\ref{g28b}) and (\ref{g28})
appears due to the limiting procedure ($\delta=1/2$ for finite
$N$, $\delta =0$ for $N=\infty$). Direct consideration of
potential well in WKB approximation gives the same result as in
(\ref{g28}).

Expression (\ref{i7}) gives the following condition $$ 1\ll n\ll
\frac1{\gamma^2\beta} $$ of WKB approximation applicability for
infinitely high potential well ($N\to \infty$). One can see that
the approximation is not valid for small $n$ (as in undeformed
case), but it becomes invalid for very large $n$.

\subsection{$-1/X^2$ potential}\label{1x2}

Let us consider the following Hamiltonian
\begin{equation}\label{g100}
H=P^2-\frac\gamma{X^2},\quad \gamma>0
\end{equation}
Only for negative energies there exist bound states. From equation
$H(X,P)=E$ we find
\begin{equation}\label{g101}
X=\frac{\sqrt\gamma}{\sqrt{P^2-E}}.
\end{equation}
Then rule (\ref{h12s}) can be rewritten as
\begin{equation}\label{g102}
4\int\limits^\infty_0 \frac{\sqrt{\gamma}dP}{\sqrt {P^2-E}(1+\beta
P^2)}=\frac{\sqrt\gamma}{\sqrt{1+E\beta}} \ln\left(
\frac{2+2\sqrt{1+E\beta}+E\beta} {2-2\sqrt{1+E\beta}+E\beta}
\right)=2\pi(n+\delta),
\end{equation}
where $\delta$ depends on boundary conditions. Equation
(\ref{g102}) can be solved for small $\beta$ and it gives
\begin{equation}\label{g103}
E_n=-\frac4\beta e^{-\pi(n+\delta)/\sqrt\gamma}.
\end{equation}
For $\beta\to0$ one can see that $E_n\to-\infty$. It corresponds
to the fact that for undeformed case there does not exist any
bound state for this potential. So, deformation of the space leads
to the existence of bound states for $-1/X^2$ potential.

\correction{For a singular potential $V(X)$ we can estimate
$P\approx\sqrt{2m|V(X)|}$ at the vicinity of the singularity
point. It is easy to show that inequality (\ref{h14}) does not
hold for potentials $-1/X^2$ and $-1/X$ either in deformed or
undeformed spaces at the vicinity of the origin. So, formally the
WKB approximation cannot be applied to $-1/X^2$ potential. On the
other hand, the Bohr-Sommerfeld quantization rule gives exact
result for $-1/X$ potential in deformed space \cite{Fityo05}. In
undeformed case the WKB approximation also can be applied to
singular potentials (see for an instance \cite{Gordeyev97}). Thus,
we may expect that obtained spectrum (\ref{g103}) are quite
accurate too.}

\section{3D examples}

In the second section we prove the Bohr-Sommerfeld quantization
rule (\ref{h12}), (\ref{h12s}) for 1D case, then in the next
section we illustrated the rule with the examples. The examples
presented in this section demonstrate that we can use
Bohr-Sommerfeld rule in 3D space and obtain satisfactory results.

Deformed commutation relation usually is generalized to the
following form in 3D case \cite{Kempf97b}:
\begin{equation}\label{g8}
[X_i,P_j]=i(1+\beta P^2)\delta_{ij}+i\beta'P_iP_j.
\end{equation}
There exists a simple momentum representation for coordinate and
momentum operators:
\begin{equation}\label{g9}
X_i=(1+\beta p^2)x_i+\beta'p_i\sum_{j=1}^3p_jx_j,\ P_i=p_i.
\end{equation}
In \correction{a} semiclassical approach coordinate and momentum
operators are substituted with corresponding variables and
\begin{equation}\label{g10}
X^2=\left[1+(\beta+\beta')p^2\right]^2x_p^2+(1+\beta
p^2)^2\frac{L^2}{p^2},
\end{equation}
here we use spherical system of coordinates $(p,\theta,\phi)$;
$x_p=\frac{(\vec x\cdot \vec p)}{p}$ denotes ``radial part'' of
coordinate, $L^2$ is an angular part. \correction{The} 3D problem
in semiclassical approach can be reduced to \correction{a} 1D one
if one substitutes $L^2$ with $\left(l+1/2\right)^2$
\cite{Landau}.

\subsection{Hydrogen atom}\label{ha}

The classical Hamiltonian reads
\begin{equation}\label{g11}
H(p,x)=P^2-\frac{\gamma}X=p^2-\frac\gamma {
\sqrt{\left[1+(\beta+\beta')p^2\right]^2x_p^2+(1+\beta
p^2)^2\left(\frac{l+1/2}{p}\right)^2} }.
\end{equation}

\correction{The} energy values of bound states of hydrogen atom
are negative. Then from the equation $H(p,x)=E$ we obtain
\begin{equation}\label{g12}
x_p=\frac{1}{1+(\beta+\beta')p^2}{\sqrt{\frac{\gamma^2}{(p^2-E)^2}
- (1+\beta p^2 )^2 \left(\frac{l+1/2}{p}\right)^2 }}
\end{equation}
and Bohr-Sommerfeld quantization condition reads
\begin{equation}\label{g13}
2\int\limits_{p_{min}}^{p_{max}}x_pdp=2\pi\left(n+\frac12\right).
\end{equation}
The integral (\ref{g13}) is very cumbersome, so we expand it in
powers of $\beta$ and $\beta'$. In linear approximation it reads
\begin{eqnarray}\label{g14}
\int\limits_{p_{min}}^{p_{max}}x_pdp\approx \int
\limits_{p_{min}}^{p_{max}} \sqrt{\ldots\mathstrut}\, dp -
(\beta+\beta') \int \limits_{p_{min}}^{p_{max}}
\sqrt{\ldots\mathstrut}\, p^2dp -\beta\left(l+\frac12\right)^2
\int \limits_{p_{min}}^{p_{max}} \frac{dp}
{\sqrt{\ldots\mathstrut}\,},
\end{eqnarray}
where
$$
\sqrt{\ldots\mathstrut}\,=\sqrt{\frac{\gamma^2}{(p^2-E)^2}-
\left(\frac{l+1/2}p\right)^2}.
$$

\correction{The} integration of (\ref{g14}) gives
\begin{equation}\label{g16}
-\pi\left(l+\frac12\right)+\frac{\gamma\pi}{2\sqrt{-E}}- \pi
(\beta+\beta')\gamma\left(\frac{\gamma}{4(l+1/2)}-\frac{\sqrt{-E}}2
\right)-\pi\beta\frac{\gamma^2}{4}\frac1{l+1/2}.
\end{equation}
Then solution of equation (\ref{g13}) in linear  approximation
gives
\begin{equation}\label{g17}
E_{n,l}\approx -\frac{\gamma^2}{4n^2}+\frac{\gamma^4}{8n^3} \left(
\beta\left[\frac2{l+1/2}-\frac1n\right]+\beta'
\left[\frac1{l+1/2}-\frac1n\right] \right).
\end{equation}

We compare this result with expression for the correction obtained
by Benczik and collaborators \cite{Benczik05} with the help of
perturbative theory. Their expression contains one additional term
$$\frac{\gamma^4}{16n^3}\frac{2\beta-\beta'}{l(l+1)(l+1/2)}.$$ For
large $l$ the term is small in comparison with the rest terms. So,
for 3D hydrogen atom Bohr-Sommerfeld quantization rule provides
satisfactory accuracy.

\subsection{Harmonic oscillator}\label{ho3}

The Hamiltonian of the system is
\begin{equation}\label{g18}
H=P^2+X^2=p^2+\left[1+(\beta+\beta')p^2\right]^2x_p^2+(1+\beta
p^2)^2\frac{(l+1/2)^2}{p^2}.
\end{equation}
From equation $H(p,x)=E$ we obtain
\begin{equation}\label{g19}
x_p=\frac1{1+(\beta+\beta')p^2}\sqrt{E-p^2-(1+\beta
p^2)^2\left(\frac{l+1/2}p\right)^2}.
\end{equation}
Corresponding contour integral can be calculated exactly but it is
very cumbersome. On the other hand, as we see Bohr-Sommerfeld
quantization rule gives correct result only in linear
approximation over $\beta$, $\beta'$. Therefore, we calculate it
in the linear approximation:
\begin{equation}\label{g20}
\frac\pi2E-\frac\pi2(\beta-\beta')(l+1/2)^2-
\pi(l+1/2)-\frac\pi8(\beta+\beta')E^2=2\pi(n_p+1/2),
\end{equation}
from which we obtain
\begin{equation}\label{g21}
E_n=2n+3+(\beta+\beta')(n+3/2)^2+(\beta-\beta')(l+1/2)^2,
\end{equation}
where $n=2n_p+l$. The spectrum of 3D harmonic oscillator was
calculated exactly in \cite{Chang02}. The difference of their
exact expression and our approximate one is
\begin{equation}\label{g22}
2\beta-\frac{\beta'}2.
\end{equation}
For large $n$ and $l$ we see the method is good again.

\section{Concluding remarks}

In the paper we derived Bohr-Sommerfeld quantization rule and
considered its applicability for 1D case. A new interesting
feature appearing in the presence of deformation is that WKB
approximation becomes valid for intermediate quantum numbers, but
it can become invalid for small (as in undeformed case) as well as
for very large quantum numbers. This feature is illustrated with
example \ref{aho}, an infinitely high potential well.

To verify the method we compared results obtained with the help of
Bohr-Sommerfeld quantization rule with exact spectra expressions
for harmonic oscillator and infinitely high potential well
(examples \ref{ho}, \ref{aho}) and show that the obtained results
are asymptotically exact for large $n$. The consideration of
$-1/X^2$ potential indicates that there may exist bound states for
this potential in deformed space, although for undeformed case
bound states do not exist (example \ref{1x2}). It was shown that
the method could be applied to 3D problems with radial symmetry
with satisfactory accuracy (see examples \ref{ha}, \ref{ho3}).

As a result it seems that the Bohr-Sommerfeld quantization rule
can be applied to consideration of a wide variety 1D problems as
well as to 3D problems in deformed space.

\section{Acknowledgment}

The authors thank Dr.\ Andrij Rovenchak for careful reading the
ma\-nu\-script.

\end{document}